\newcommand\Tab[1]{Table~\ref{tab:#1}}
\newcommand{\be}{\begin{equation}}
\newcommand{\ee}{\end{equation}}
\newcommand\beq{\begin{eqnarray}}
\newcommand\eeq{\end{eqnarray}} 
\newcommand\eqn[1]{\label{eq:#1}} 
\newcommand\eq[1]{eq. (\ref{eq:#1})}

\newcommand{\vev}[1]{\langle #1 \rangle}

\newcommand{\CE}{{\cal E}}

\newcommand{\CN}{{\cal N}}

\newcommand\half{{\textstyle{\frac{1}{2}}}}

\newcommand{\mybar}[1]%
        {\kern 0.6pt\overline{\kern -0.6pt#1\kern -0.6pt}\kern 0.6pt}
\documentclass[cits]{PoS}

\title{Listening to Noise}

\ShortTitle{Listening to Noise}

\author{Michael G. Endres\\
        Theoretical Research Division, RIKEN Nishina Center, Wako, Saitama 351-0198, Japan\\
        E-mail: \email{endres@riken.jp}}

\author{\speaker{David B. Kaplan}\\
        Institute for Nuclear Theory, University of Washington, Seattle, WA 98195-1550, USA\\
        E-mail: \email{dbkaplan@uw.edu}}
        
        \author{Jong-Wan Lee\\
        Institute of Particle and Nuclear Studies (IPNS), KEK, 1-1 Oho, Tsukuba, Ibaraki, Japan\\
        E-mail: \email{jongwan@post.kek.jp}}
        
        \author{Amy N. Nicholson\\
        Deptartment of Physics, University of Maryland, College Park MD 20742-4111, USA
\\
        E-mail: \email{amynn@umd.edu}}

\abstract{We show how sign problems in simulations of many-body systems can manifest themselves in the form of heavy-tailed correlator distributions, similar to what is seen in electron propagation through disordered media.  We propose an alternative statistical approach for extracting ground state energies in such systems, illustrating the method with a toy model and with lattice data for unitary fermions.
}

\FullConference{ The XXIX International Symposium on Lattice Field Theory - Lattice 2011\\
July 10-16, 2011\\
Squaw Valley, Lake Tahoe, California}

\begin{document}

\section{Introduction}

While QCD is accepted as the correct theory of the strong interactions, it is frustrating that we still are unable to derive from it the basic properties of matter or map out a phase diagram that is more than a cartoon.  The problem is that the only tool we have for analyzing QCD nonperturbatively outside of special kinematic regions is the lattice, and lattice QCD faces daunting barriers to investigating systems with significant baryon number.  In the grand canonical formulation, the problem encountered is that the fermion determinant has a rapidly varying phase leading to strong cancelations in the path integral and a suppression of the partition function that is exponential in the volume.  This is often called the ``fermion sign problem", although we will argue it is not specific to  fermions. In a canonical formulation, where one computes correlation functions of fixed baryon number,  one encounters a couple of problems.  The first is that the measurements may become increasingly noisy as the baryon number is increased, which we will call a ``noise problem".  The second is that the value obtained for the correlator may be relatively quiet, yet drift with time, giving no evidence for a plateau in the effective mass from which one can infer a ground state energy.  The effective mass computed from a correlator in a system with discrete nonzero energy levels must exhibit a plateau at late time;  so if a plateau is absent  it follows that the  path integral is not being estimated correctly, with little overlap between the field configurations being sampled and the ones 
that actually dominate  in a correct estimation of the correlator.  We will refer to this second problem as an ``overlap problem". In an actual simulation both problems may be found simultaneously: a drifting and noisy value for the correlator.

After arguing that the problems encountered in the canonical and grand canonical approaches arise from the same physics, we focus on the canonical method and the statistical distribution of correlators.  We give examples where the overlap problem is a manifestation of an underlying statistical distribution for the correlator which is very heavy-tailed, and not too far off a log-normal distribution.  We show that the appearance of nearly log-normal distributions is indicative of a mean field theory expansion, such as that used to study electron properties in disordered media. We then develop a technique to effectively overcome the overlap problem by defining mass plots using an  estimator which is not minimum bias, but where the log-normal result is the first term in  a well-defined expansion.  This method has been used with great success in studies of trapped ``unitary fermions" --- a strongly coupled nonrelativistic conformal theory relevant to trapped atoms at a Feshbach resonance --- allowing a measurement of the ground state energy of up to 70 fermions in a harmonic trap.  We also show QCD data from the NPLQCD collaboration that indicates similar techniques might aid  in the study of QCD; we expect this to be especially so at larger baryon number.  This work has been published in Ref.~\cite{endres2011noise}.

As a possible direction for future research, our approach suggests a renormalization group approach to the statistical distributions of lattice observables, and a way to analyze such distributions in a manner analogous to effective field theory, where the $n^{th}$  cumulants of the log of the observable play the role of  higher dimension operators in the effective Lagrangian.

\section{Noise and the physical spectrum}

We first  examine the connections between the noise problem encountered in canonical simulations and the phase problem due to the fermion determinant in grand canonical computations, developing a heuristic picture for why a sign/noise/overlap problem arises in the first place.

Grand canonical simulations of lattice QCD at nonzero quark chemical potential involve a path integral measure that includes a fermion determinant which is neither  real nor positive, leading to an exponentially hard computation \cite{Troyer:2004ge}.  It was noticed long ago \cite{Gibbs:1986xg} that the determinant phase problem sets in for quark chemical potential $\mu \ge m_\pi/2$, even at zero temperature, which seems peculiar since nothing physical happens until $\mu>m_N/3$, $m_N$ being  the nucleon mass.  A beautiful explanation  for this precocious annoyance was given by Splittorff and Verbaarschott  \cite{Splittorff:2006fu} (see also \cite{Cohen:2004qp} for earlier related ideas):  they pointed out that in two flavor QCD with degenerate quarks, the quark determinant phase is exactly what sets apart a simulation at finite isospin chemical potential ($\mu_u=-\mu_d=\mu$)  from a simulation at finite quark number ($\mu_u=\mu_d=\mu$).  In the former system, pion condensation occurs at $\mu=m_\pi/2$ leading to a drop in the free energy density, and therefore growth of the partition function $Z$ which is exponential in the volume.  Therefore a simulation at finite baryon number with   $\mu\ge m_\pi/2$ must have the quark determinant phase oscillate wildly to cancel this exponential growth, since there is no pion condensate in this system.  The sign problem in QCD is therefore characterized by $\Delta\mu = \zeta\equiv(M_N/3 - m_\pi/2)$, the difference between the values for $\mu$ where the ground states rearrange themselves in the isospin and baryon number systems.  

\begin{figure}[t]
\centerline{\includegraphics[width=7 cm]{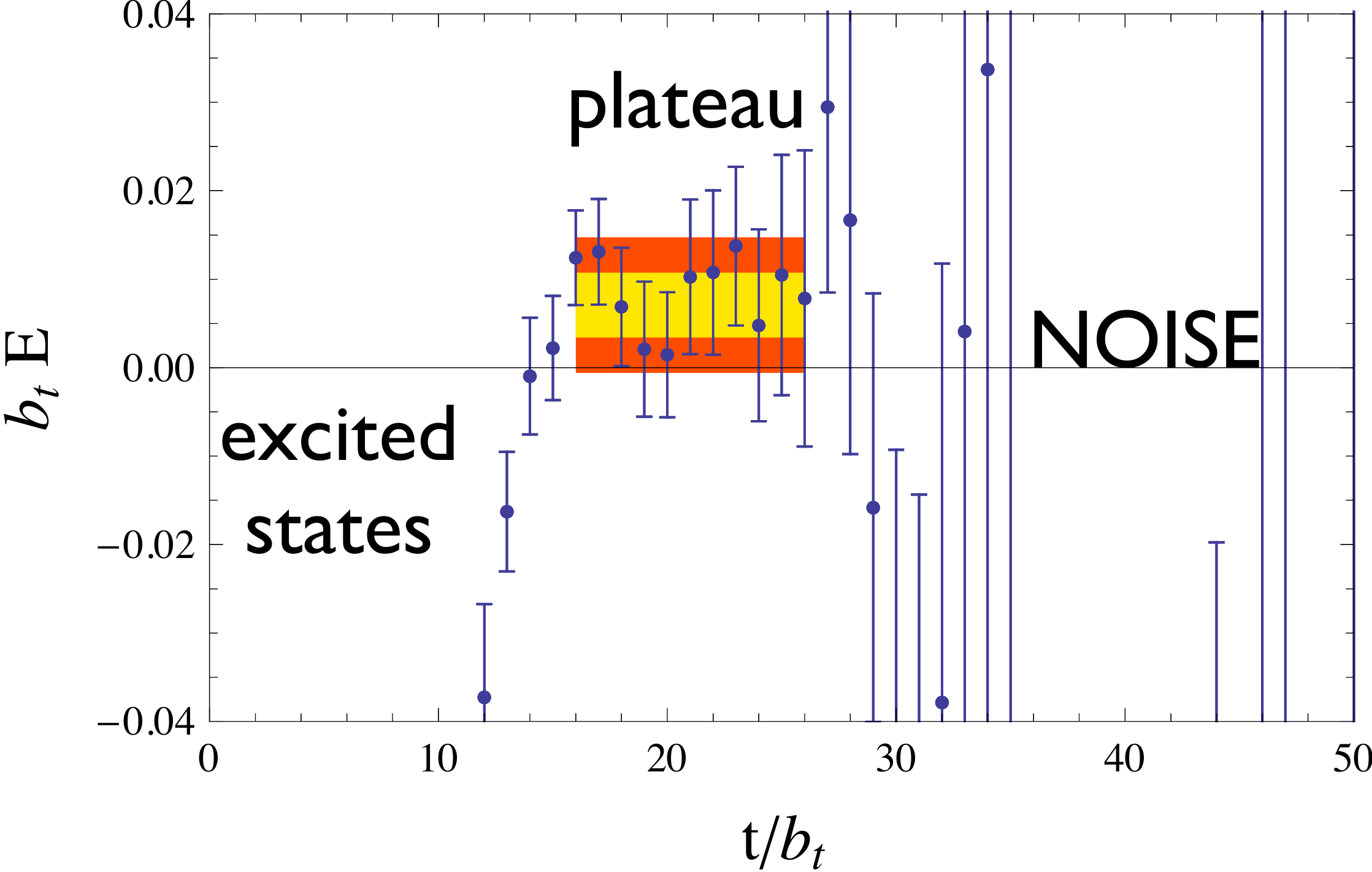}}
\caption{\label{fig:triton} {Effective mass plot for nine quarks in the triton channel from NPLQCD \cite{Beane:2009gs}; annotation by D.K.
 } }
\end{figure}

In a canonical simulation  of the $A$-nucleon state  one looks instead at the $3A$ quark correlator over Euclidian time $\tau$, $C_A(\tau,U)$, which is a function of the link fields $U$ for the gluons.  For short times, the correlator gets contributions from the many excited states with $3A$ quarks, but at later time it should be dominated by the ground state for   $A$ nucleons.  One typically creates an effective mass plot of 
\beq
m_{\rm eff} (\tau)=- \frac{1}{\Delta\tau}\ln \vev{C_A(\tau+\Delta\tau,U)}/\vev{C_A(\tau,U)}
\eeq
where the average is over an ensemble of link fields and $\Delta\tau=O(1)$, chosen to optimize the result.  When the correlator is dominated by the ground state, the effective mass should be a constant, the ground state energy.   Fig.~\ref{fig:triton} shows an effective mass plot
for the case for $A=3$ provided by the NPLQCD collaboration clearly exhibiting three types of behavior: strong $\tau$-dependence at short times when excited states contribute, a plateau dominated by the mass of the triton, and degradation into noise at later time. 

\begin{figure}[b]
\centerline{\includegraphics[width=7cm]{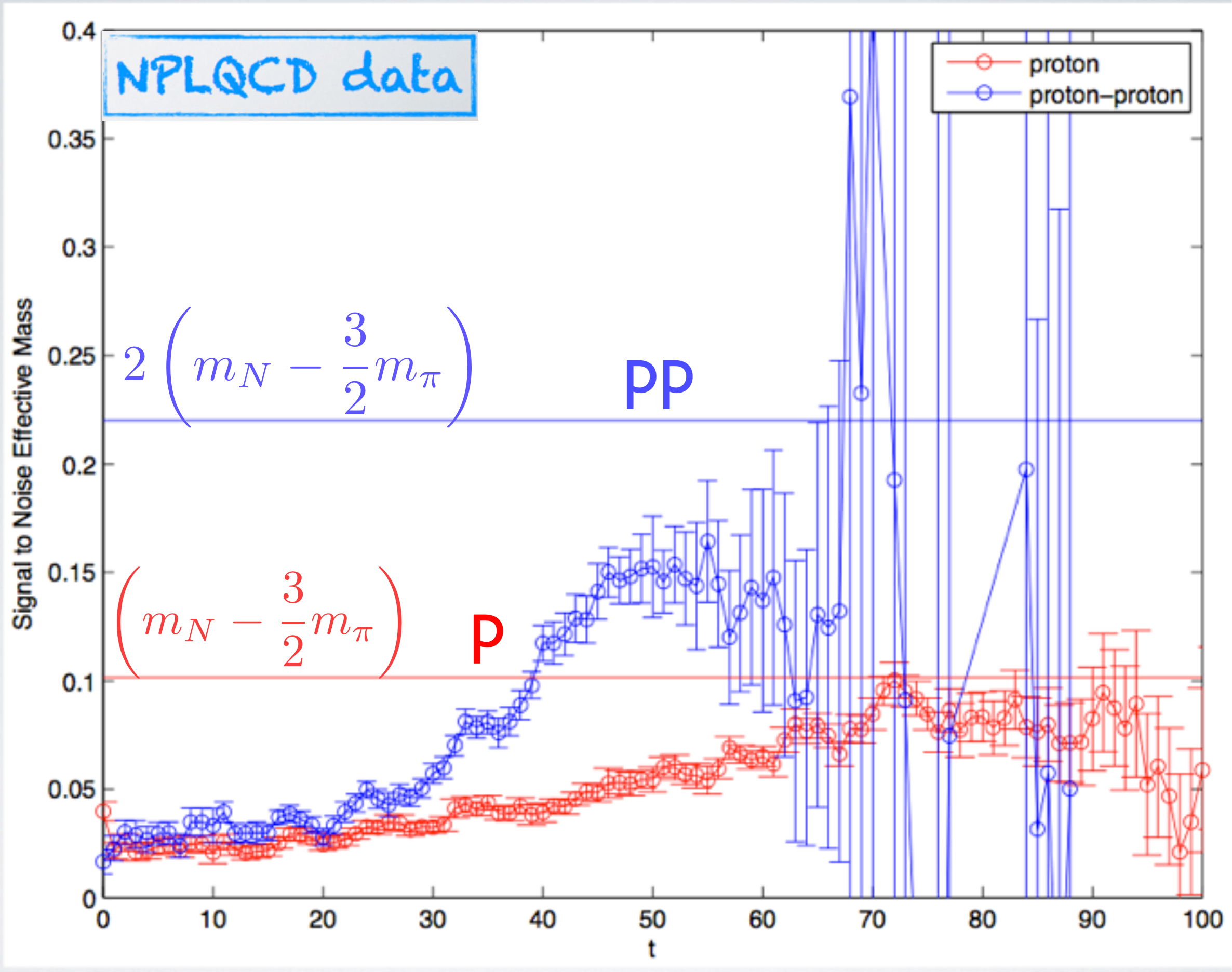}}
\caption{\label{fig:lepage} {A plot of $-1/t \ln$(signal/noise ratio) versus time for $p$ (red) and $pp$ (purple) correlators computed in lattice QCD; horizontal lines give the expected values for these quantities using the Lepage argument.  Figure supplied by K. Orginos (NPLQCD), annotation by D.K.
 } }
\end{figure}

The growth of noise in the correlator makes it difficult to extract a signal without a lot of statistics, and the problem is expected to get worse with increasing baryon number. It would have been strange if properties of matter at finite baryon density, which are extremely difficult to compute in the grand canonical ensemble, were easy to compute in a canonical simulation --- and indeed this noise problem exhibited by the correlator at late time is presumably an avatar of the sign problem.  
This sign/noise problem does not arise simply because of Fermi statistics; for example, constructing correlators $C_A$ as $3A\times 3A$ Slater determinants of quark propagators accounts for Fermi statistics and leads to a computational cost from the determinant   only scaling as $(3A)^3$, not the exponential difficulty seen in the grand canonical computation. 
Instead, the  noise is closely related to the  physical spectrum, as  has been quantified by Lepage    \cite{Lepage:1989hd}.
For example, in QCD the expectation of a $3A$ quark correlator for a nucleus  of atomic number $A$ and mass $M_A$  is $\langle C_A\rangle\sim e^{- M_A \tau}$, while the  variance in the sample mean $\mybar C_A$  can be estimated as 
\beq
\sigma^2= \frac{1}{\CN}\left(\langle C_A^\dagger C_A\rangle -\langle C_A^\dagger\rangle\langle C_A\rangle \right)\sim \frac{1}{\CN} e^{-3A m_\pi \tau}
\eeq 
for sample size $\CN$.
Since $C_A$ corresponds to $3A$ quark propagators and $C_A^\dagger$ to $3A$ anti-quark propagators, the variance is dominated by the state with $3A$ pions.   Thus the signal  to noise ratio scales as $\sim \sqrt{\CN}\,\exp(-3A\zeta \tau)$, where $\zeta=(M_N/3- m_\pi/2)$ is the same parameter we saw characterizing the sign problem in the grand canonical case.  This reasoning is rather simplistic, as the overlap between the operators and the nucleon or pion states --- which will typically contain volume factors -- has not been included.  However, Fig.~\ref{fig:lepage} shows evidence from QCD simulations by NPLQCD that the Lepage argument is qualitatively correct.  The noise and sign problems are therefore presumably closely related and determined by the physical spectrum of the theory and should not be thought of as solely a ``fermion sign problem"; similar issues can also arise in interacting boson theories.  

In either case the problem  arises from the existence of multiparticle states for which the energy/constituent is lower than for the states one wants to study:  in QCD, quarks are in some sense lighter when they are in a pion than when they are in a nucleon, by the amount  $\zeta$.  This is a problem that can occur in bosonic systems as well, and so one should not expect the sign problem to simply be a fermion problem. Simple bosonic path integrals in the grand canonical approach, such as gauged theories with a $\phi^4$ interaction, manifestly have complex action in Euclidian spacetime and a sign problem.

Applying the Lepage argument for $C_A(\tau,U)$ to higher moments one sees that all odd moments are proportional to $\exp(-M_A\tau)$
while even moments are all dominated by pion states\footnote{Thanks to Martin Savage for first pointing this out to us.}.  Thus at very late time $\tau$, the probability distribution for $C_A(\tau,U)$ will be almost symmetric, implying a noisy sample with small mean.  At earlier times, however, we will see that the distribution appears to be very asymmetric and heavy tailed in both QCD and other theories, and it is this regime we wish to analyze.

\section{Unitary fermions and a Mean Field Description} 
 \label{sec:mf}
 
Nonrelativistic spin half fermions with strong short-range interactions tuned to a conformal fixed point where the phase shift satisfies $\delta(k)=\pi/2$ for all $k$  are called ``unitary fermions".  One can think of the inter-particle interaction as being described by a square well potential tuned to have a single bound state with zero binding energy, as one takes the width of the potential to zero. The   field theory describing this system is conformal, and interesting to study both for its simplicity and universality, its challenges for many-body theory, and because it can be realized and studied experimentally using trapped atoms tuned to a Feshbach resonance \cite{PhysRevLett.92.090402}. It is also an ideal  nonperturbative theory for studying fermion sign problems on the lattice, being simpler and faster to simulate than QCD.   At its most basic, the lattice action involves the simplest discretization of the Euclidean Lagrangian \cite{Chen:2003vy}
\beq
\psi^\dagger (\partial_\tau -\nabla^2/2M)\psi -\half m^2 \phi^2 + \phi \psi^\dagger\psi
\eqn{lag}
\eeq
 where $\phi$ is a non-propagating auxiliary field with $m^2$  tuned to a critical value, and $\psi$ is a spin $\half$ fermion with mass $M$; a more sophisticated action tuned to reduce discretization errors was recently presented in \cite{Endres:2010sq,unitary:2011aa}.
 
\begin{figure}[t]
\centerline{\includegraphics[width=7 cm]{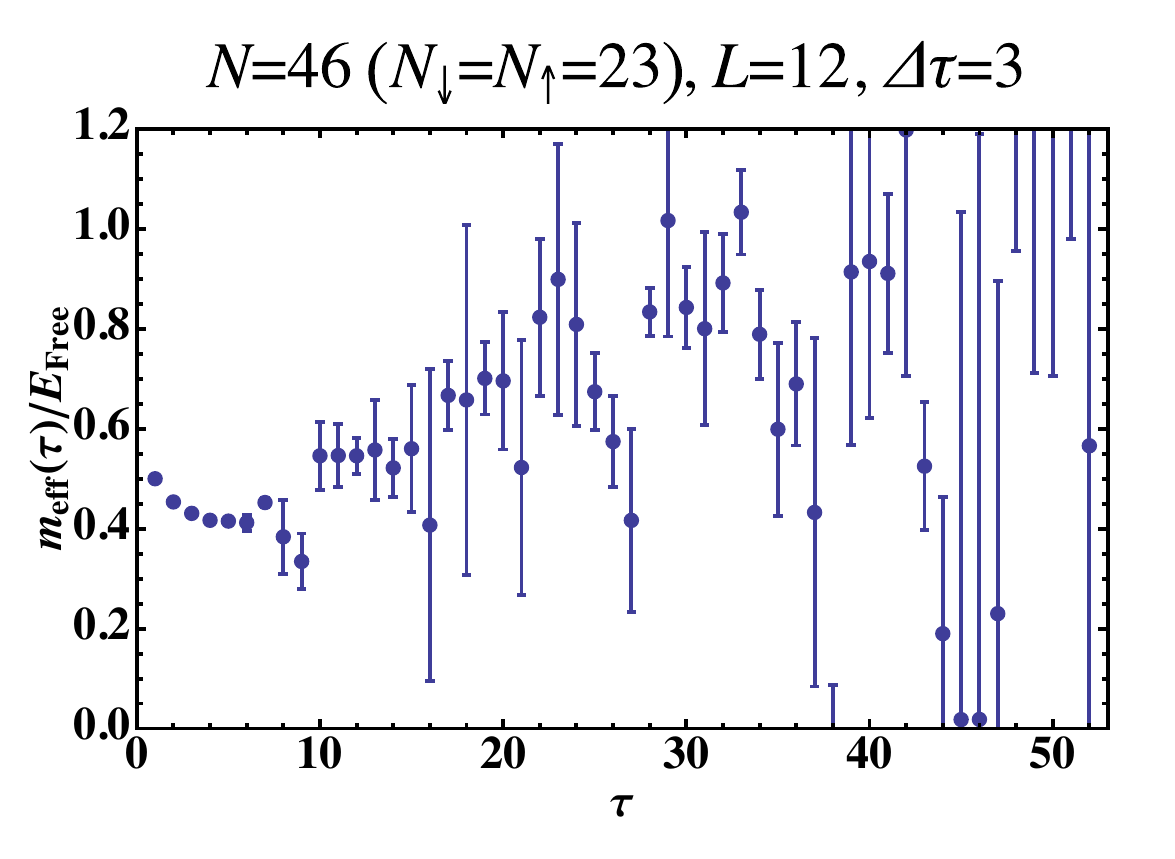}}
\caption{ The effective mass plot for 46 unitary fermions in a box of size $L=12$.
} 
\label{fig:N46noise}
\end{figure}

\begin{figure*}[b]
\begin{tabular}{cc}
\includegraphics[width=7 cm]{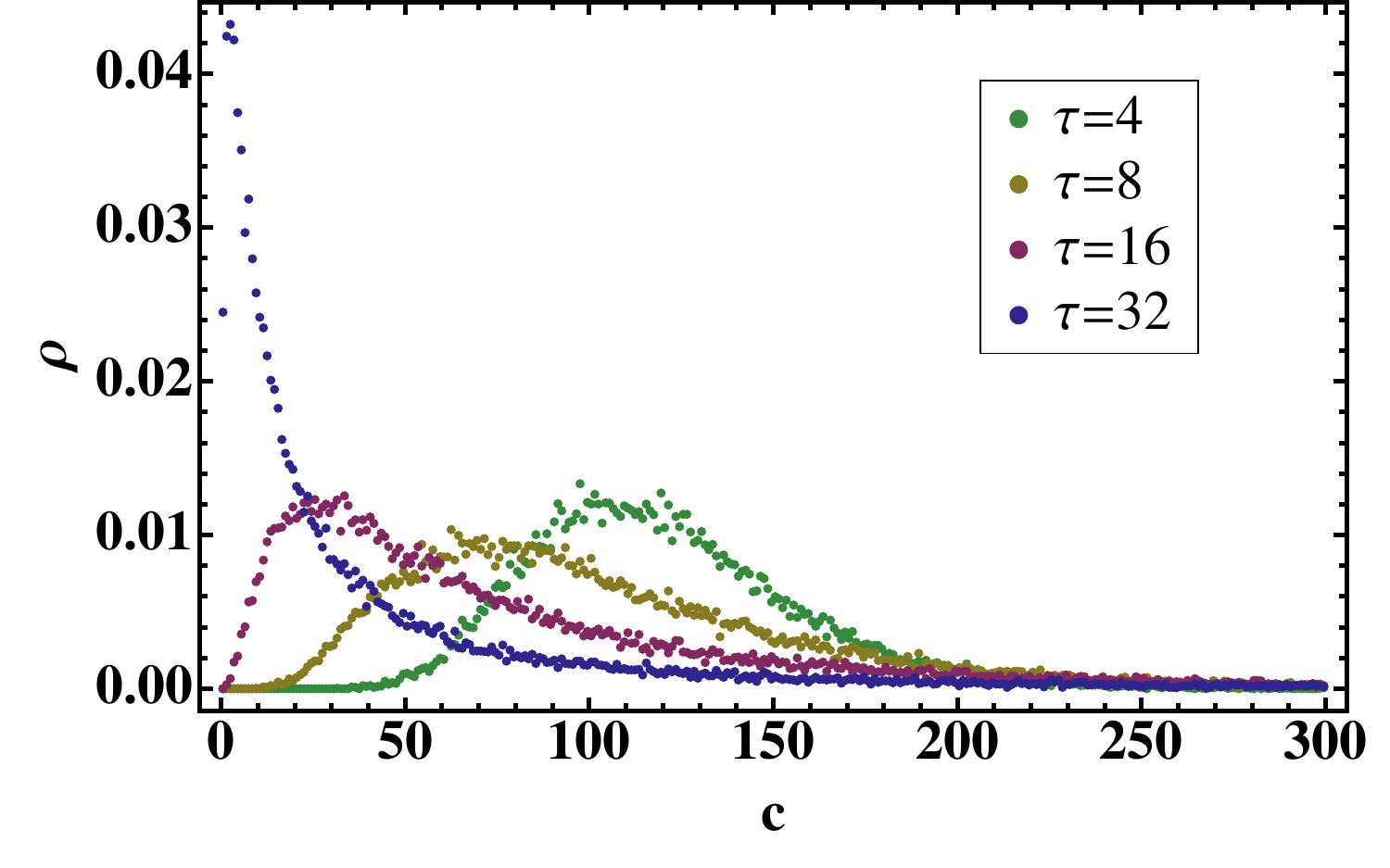}
\includegraphics[width=7 cm]{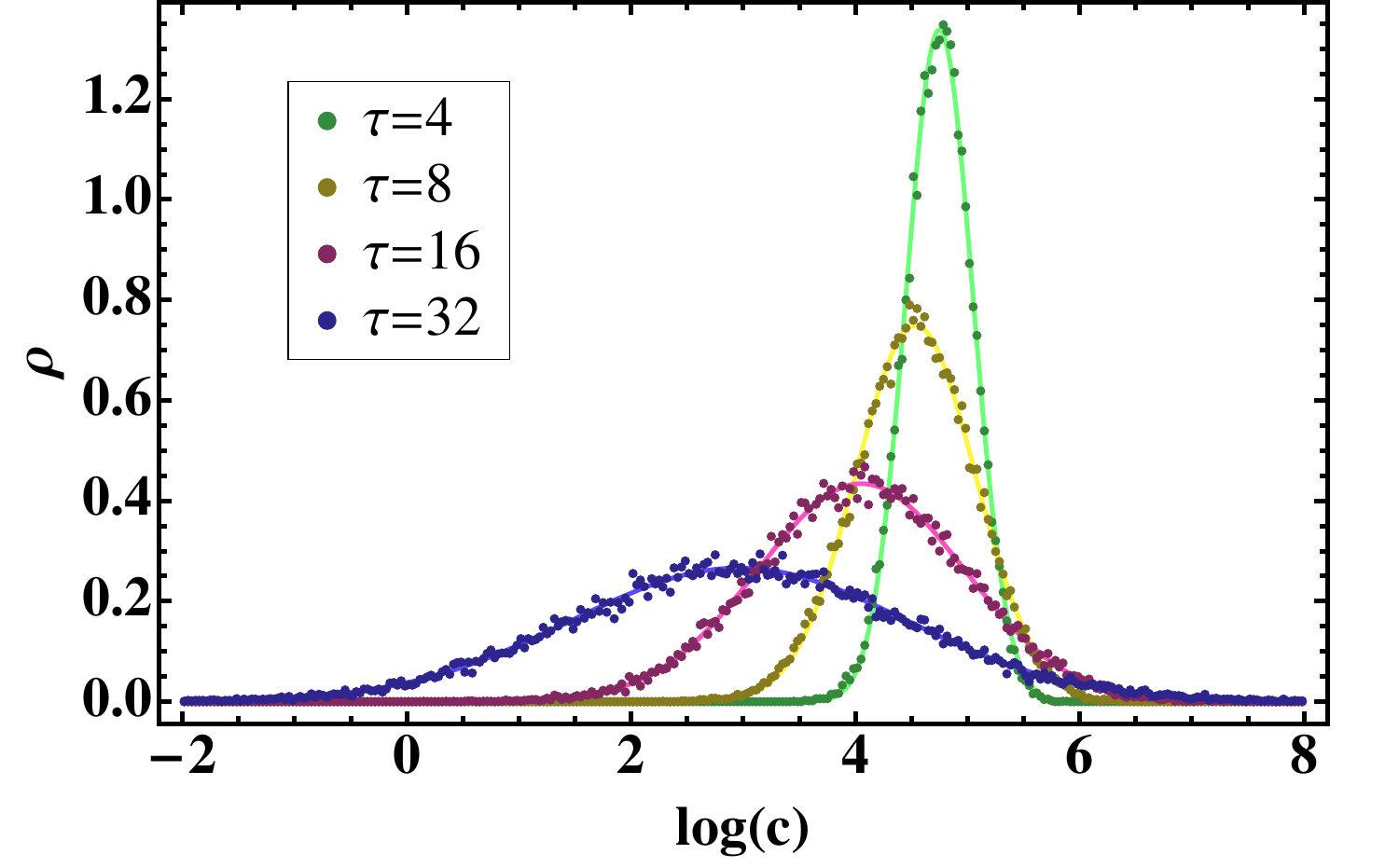}
\end{tabular}%
\caption{\label{fig:ln} {Distribution histograms for $c=C_N(\tau,\phi)$ and $\ln (c)$ for $N=4$ unitary fermions at several times $\tau$, taken from Ref.~\cite{unitary:2011aa}.  Curves fitting $\ln(c)$ are Gaussian, implying that $c$ is approximately log-normal distributed, with $\sigma^2$ increasing with time.  } }
\end{figure*}

 The ground state for $N=(N_{\uparrow}+N_{\downarrow})$ unitary fermions can be determined by computing the correlator for each background $\phi$ field, and then averaging over the ensemble of such fields.  One finds that when $N$ is large, the correlators rapidly become extremely noisy and drift away from a plateau; see for example Fig.~\ref{fig:N46noise} for an effective mass plot for $N=46$ unpolarized fermions in a box with periodic boundary conditions of size $L=12$.  In order to better understand the problems apparent for $\tau \gtrsim 10$, it is instructive to plot a histogram of the correlator before averaging over the $\phi$ field.   This is shown in   Fig.~\ref{fig:ln}, which
reveals a distribution for $N$-body correlators $C_N(\tau,\phi)$ which is increasingly non-Gaussian at late $\tau$; in fact, Fig.~\ref{fig:ln} shows that it is $\ln\, C_N$ which appears to be roughly normally distributed, so that  $C_N(\tau,\phi)$  is approximately {\it log-normal} distributed with an increasingly large $\sigma$ and long tail at late time.  

A log-normal distribution for the correlator is difficult to reconcile with the Lepage argument.  First of all, Lepage assumes that the central limit theorem applies and that the distribution of correlators  is close to Gaussian  and well described by the mean and variance, which is evidently not the case here.  As we will see in a toy model below, the Central Limit Theorem may be irrelevant in such cases, requiring an exponentially large number of configurations to be applicable. Secondly, the size of the variance for this log-normal distribution cannot be related to the physical spectrum of the theory using the Lepage argument.  The Lepage argument relates the variance in measurements of $C_N$ to the energy of the lightest state coupling to $(C_N)^2$; there are no anti-particles in this theory, and  and so $(C_N)^2$ couples to the ground state  of a system  $N$ particles each of two different species of spin $1/2$ unitary fermions \footnote{If $C_N$ is the  product of $N_\downarrow$ antisymmetrized correlators times $N_\uparrow$ antisymmetrized correlators, then $(C_N)^2$ corresponds to a product of four groups of antisymmetrized  correlators, and  is equivalent to  the propagator for  two distinguishable species of spin half fermions.}.  However, such a theory in the continuum does not have a finite energy ground state, being unstable against condensing into a point-like clump.  On the lattice, a ground state exists for such a system since there is a maximum density due to Fermi statistics and the cutoff;  this ground state is a lattice artifact, however, and even then may have a vanishingly small probability to be created by  $(C_N)^2$, in which case the variance might be determined by some metastable or excited state.

\begin{figure}[b]
\centerline{\includegraphics[width=7 cm]{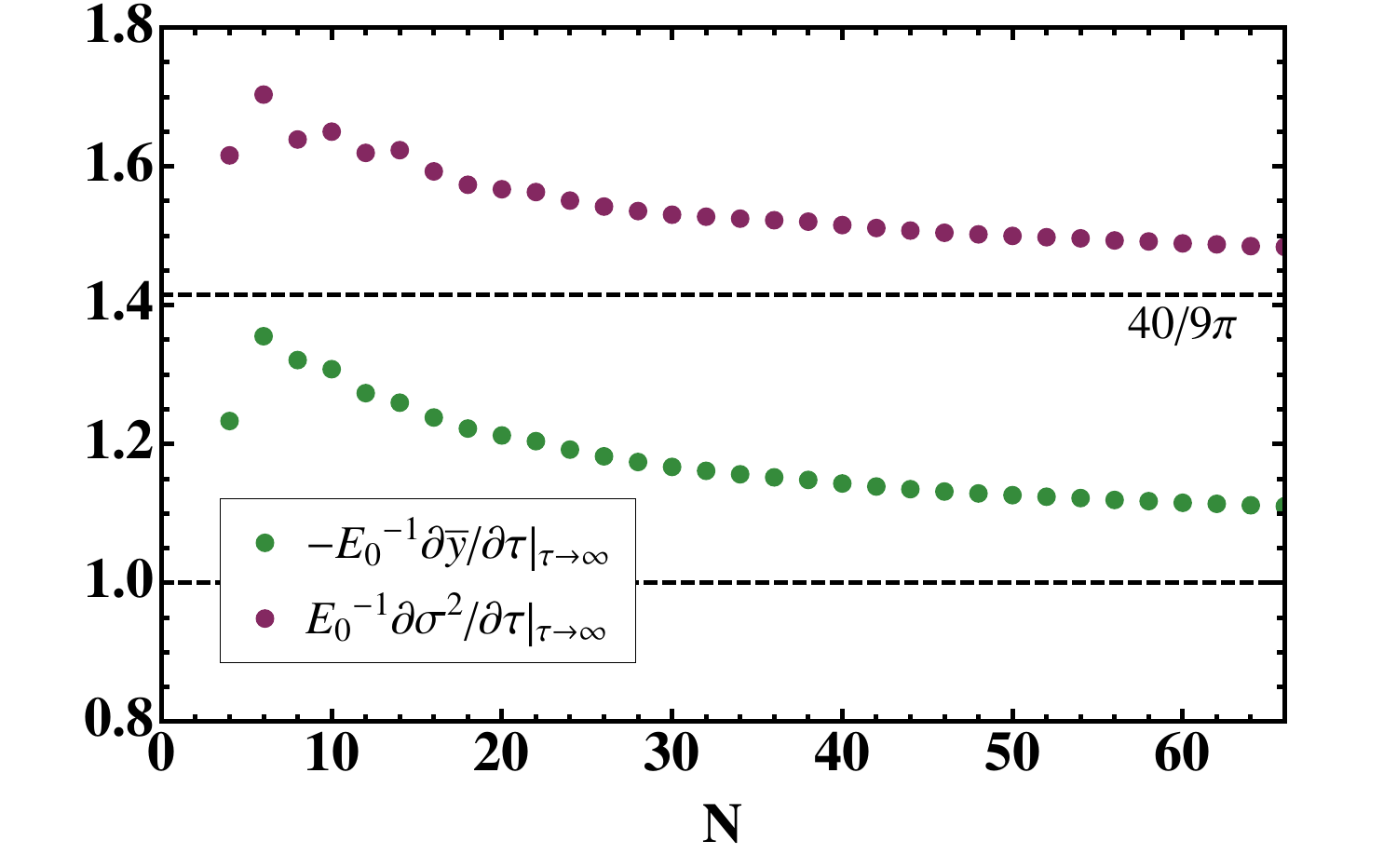}}
\caption{The quantities $-\frac{1}{E_0}\frac{\partial\mybar y}{\partial\tau}$ and $\frac{1}{E_0}\frac{\partial\sigma^2}{\partial\tau}$ as a function of $N$ for unitary fermions at late times on a lattice of size $L=10$, compared to mean field prediction 
(dashed lines).}
\label{fig:sigmu}
\end{figure}

The appearance of a heavy-tailed distribution should not be surprising, however, since having fermions wander through a random $\phi$ background is similar to the problem of  electrons propagating in disordered media, where  heavy-tailed distributions are ubiquitous  in the vicinity of  the Anderson localization transition. For example, it is found that for physical quantities such as the current relaxation time or normalized local density of states, the distribution function $P(z)$ scales as $\exp(-C_d \ln^dz)$. A particularly simple way to derive these results is to use the optimal fluctuation method of Ref.~\cite{smolyarenko1997statistics}, which is a mean field approach.  We can adapt these methods to the current problem, defining the variable $Y=\ln C_N(\tau,\phi)$ and computing its probability distribution $P(y)$ as
\beq
P(y) \propto   \int D\phi e^{-S_\phi}\, \delta(Y(\tau,\phi)-y)= \int D\phi\,\frac{dt}{2\pi} e^{-S}
\eeq
where $S_\phi = \int d^4x\, \textstyle{\frac{m^2}{2}\phi^2}$ and  $S= S_\phi -it(\ln C_N(\tau,\phi)-y)$. Using the PDS subtraction scheme \cite{Kaplan:1998tg}
we have $m^2 = M\lambda/4\pi$, where the renormalization scale $\lambda$ is taken to be the physical momentum scale in the problem --- in this case  $\lambda = k_F \equiv (3\pi^2N/V)^{1/3}$, $N/2$ being the number of fermions with a single spin orientation. We proceed now to evaluate this integral using a mean field expansion; it is not evident that there is a small parameter to justify this expansion, but the leading order  result is illuminating and fits the  numerical data well.  We expand about $\phi(x)=\phi_0$, $t=t_0$, and use the fact that  for large $\tau$ the $n^{th}$ functional derivative of $\ln C_N(\tau,\phi)$ with respect to $\phi(x)$  equals the the  1-loop Feynman diagram with $n$ insertions of $\psi^\dagger\psi$ in the presence of a chemical potential $\mu=k_F^2/(2 M)$.  The equations for $\phi_0$ and $t_0$ are given by
\beq
t_0 &=& -i \frac{m^2\phi_0}{\vev{n(x)}_c}=-i\frac{V m^2\phi_0}{N}\cr
\phi_0 &=&- \frac{y-\ln Z + \tau E_0(N)}{N\tau}
\eeq
where $E_0(N) = 3N E_F/5 $ is the total energy of $N$ free degenerate fermions ($N/2$ of each spin), and $Z$ is the overlap of the source and sink with the free fermion state. The leading term in the mean field expansion for $P(y)$ can therefore be expressed as
$P(y)\propto \exp\left[-\frac{(y-\mybar y)^2}{2\sigma^2}\right]
$
with
\beq
\mybar y = \ln Z -\tau E_0(N)\ ,\quad \sigma^2 = \frac{40}{9\pi} E_0(N)\,\tau\ .
\eqn{mf}
\eeq
This describes a log-normal distribution for the $N$-fermion propagator $C_N(\tau,\phi)$, with both mean and variance growing in magnitude with time in units of the energy of $N$ free degenerate fermions. In Fig.~\ref{fig:sigmu} we plot the quantities $-\frac{1}{E_0}\frac{\partial\mybar y}{\partial\tau}$ and $\frac{1}{E_0}\frac{\partial\sigma^2}{\partial\tau}$ as a function of $N$ obtained from correlator distribution data for unitary fermions at late $\tau$, and find that the gross features of the results are compatible with the mean field estimates of unity and $40/9\pi$ obtained from \eq{mf}.

 \section{A toy model}

It would be useful to devise an algorithm to reliably estimate energies without having to exhaustively sample the long tail of the correlator distribution, yet without making incorrect assumptions about the exact functional form of that tail.  An approach we suggest here is to exploit the general relationship between stochastic variables $X$ and $Y=\ln X$:
\beq
\ln\vev{X} = \sum_{n=1}^\infty\frac{\kappa_n}{n!}
\eqn{alg}
\eeq
where $\kappa_n$ is the $n^{th}$ cumulant of $Y$.  This relation can be proved by noting that the generating function for the $\kappa_n$ is $\ln \phi_Y(t)$ where $\phi_Y(t)=\vev{e^{Y t}}= \vev{X^t}$ is the moment generating function for $Y$, and evaluating at $t=1$,  assumed to be  within the radius of convergence.  The motivation for investigating \eq{alg} is that if the distribution $P(X)$ were exactly log-normal, the above sum would end after the second term, as $\kappa_{n>2}$ would all vanish; therefore by replacing the $\kappa_n$ by sample cumulants $\mybar\kappa_n$ and truncating the sum at $n=n_{max}$, one might hope to have a reliable estimator for $ \ln\vev{X} $, provided that the systematic error from truncating \eq{alg} and the statistical error from sampling $\mybar\kappa_n$ can be simultaneously minimized.

Distributions with log-normal-like tails arise naturally in products of stochastic variables.  The propagator $C_N(\tau,\phi)$ for unitary fermions can be expressed in a transfer matrix formalism as the  product of $\tau$ matrices --- one per time hop --- each of which is the direct product of $N$ $V\times V$ matrices  of the form $e^{-K/2} (1+ g  \varphi) e^{-K/2}$,  where $K$ is a constant matrix (the spatial kinetic operator), $ \varphi$ is a random  diagonal matrix with $O(1)$ entries corresponding to stochastic $\phi$ fields living on the time links, and $g$ is a coupling constant (identified with $1/m^2$ from \eq{lag}) that has been tuned to a particular critical value that is $O(1)$.  Unfortunately, little seems to be known about products of random matrices beyond dimension two \cite{PhysRevE.66.066124}.  Therefore we analyze instead a toy model where we define a ``correlator" $C_\tau$  as a product of random numbers, and an ``energy" $\CE = \lim_{\tau\to\infty} \CE_\tau$  where:
\beq
C_\tau= \prod_{i=1}^\tau (1+ g \varphi_i)\ ,\quad \CE_\tau= -\frac{1}{\tau} \ln\vev{C_\tau}
\eqn{model}\eeq
where $0\le g\le 1$ and the $ \varphi_i$ are independent and identically distributed random numbers with a uniform distribution on the interval $[-1,1]$.  The exact value for the energy is obviously $\CE_\tau=0$  since the statistical average of the correlator is $\vev{C_\tau}=1$. 
The cumulants of the variable $Y=\ln (C_\tau)$ are given by
\beq
\kappa_1  &=&\tau \left [\textstyle{ \frac{1}{2}} \log \left(1-g^2\right)+\textstyle{\frac{\tanh ^{-1}(g)}{g}}-1\right]\ ,\cr
\frac{\kappa_{n}}{n!} &=&\tau\left(\textstyle{ \frac{(-1)^n}{n}}- {Li}_{1-n}\left(\textstyle{\frac{1+g}{1-g}}\right)\textstyle{ \frac{\left(2\tanh ^{-1}(g)\right)^n}{n!}}\right)\nonumber
\eeq
for $n\ge 2$; for $g<1$ one finds that the $\kappa_n/n! $ rapidly decrease as $n$ increases.

\begin{figure}[t]
\centerline{\includegraphics[width=9.0 cm]{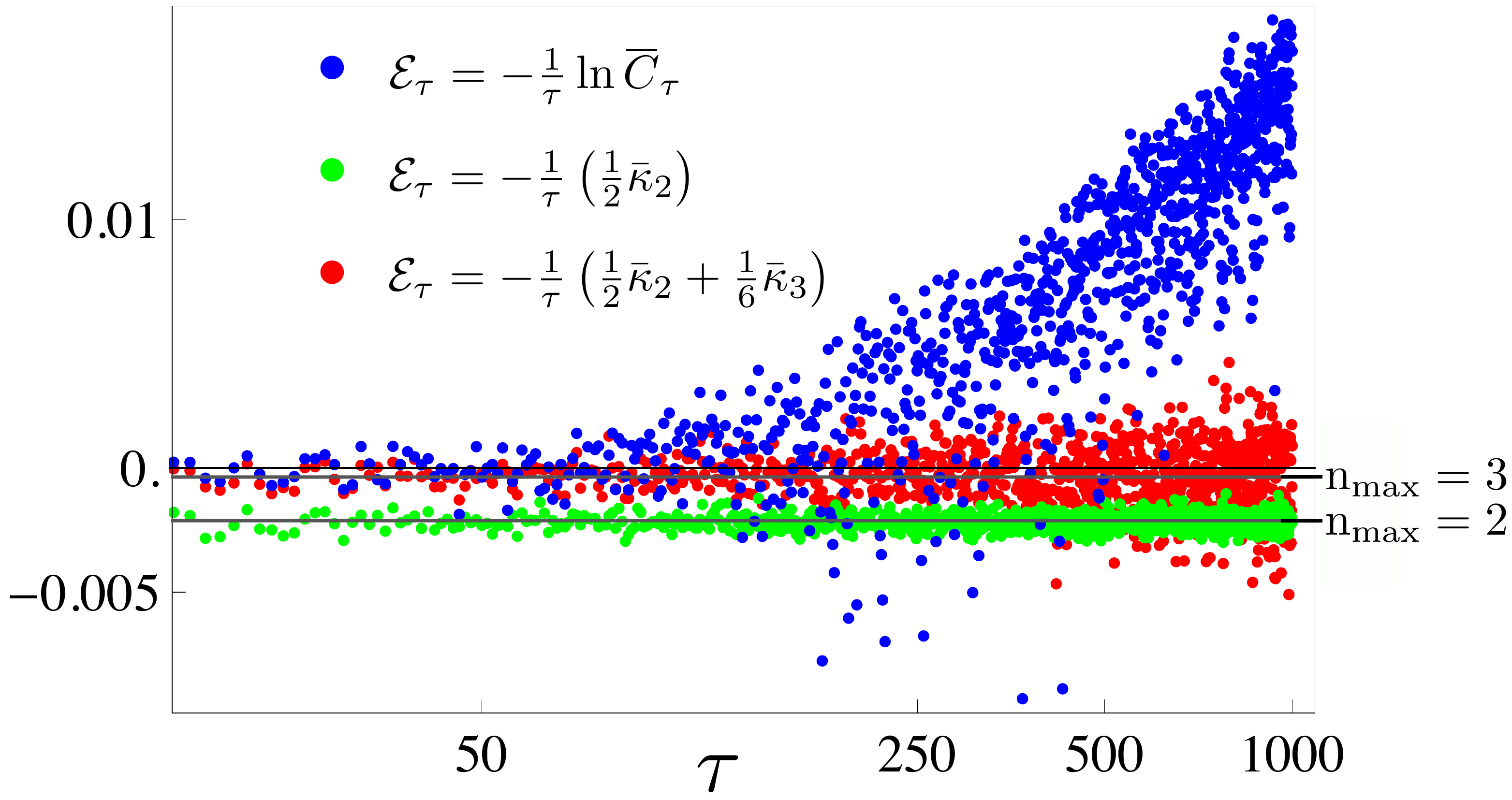}}
\caption{Simulation of the energy $\CE_\tau$ for the toy model 
eq. (4.2)
with $g=\half$.   The exact answer is $\CE_\tau=0$ (black line);    exact values of 
the expansion eq. (4.1) 
truncated at order $n_{\rm max}=2,3$ are indicated.} 
\label{fig:sim}
\end{figure}

\begin{table}[t]
\centerline{\begin{tabular}{crrr}
Method & $\CE\quad\ $ & stat. error & syst. error \\
\hline
conventional & 0.014932 & 0.002485 & -- \\
$\kappa_{n\le 2}$ &-0.002159& 0.000304& -0.002165\\
$\kappa_{n\le 3}$&-0.000412 & 0.001618&-0.000324 \\
$\kappa_{n\le 4}$&-0.000647 & 0.008379&0.000050 \\
$\kappa_{n\le 5}$&-0.001794 & 0.037561&$3.34\times10^{-6}$ \\
$\kappa_{n\le 6}$&0.010943 & 0.147739&$-1.22\times10^{-6}$ \\
 \end{tabular}}
\caption{\label{tab:table1} $\CE$ determined from 250 blocks of 50,000 configurations each for the model 
with $\tau=1000$, $g=1/2$.
}
\end{table}

 In Fig.~\ref{fig:sim} we show the results of a simulation where we compute $\CE_\tau$ for $g=\half$ and $\tau=1,\ldots,1000$.  At each value of $\tau$ we independently generated an ensemble of values for $C_\tau$ of size $N=50,000$.  From that ensemble we computed $\CE_\tau$ by 
 (i)  using the conventional estimator $\CE_\tau = -\frac{1}{\tau} \ln \mybar C_\tau$ (blue), which shows a striking systematic error for $\tau\gtrsim 50$, and statistical noise increasing up to $\tau\simeq 500$ but decreasing beyond that;
  (ii) using \eq{alg} truncated at $n_{max}=2$ using sample cumulants $\mybar\kappa_n$ (green), showing a $\tau$-independent systematic error with smaller but slowly  growing statistical error; (iii)  \eq{alg} truncated at $n_{max}=3$ (red) with a negligible constant systematic error but a larger statistical error.  Evidently, one trades systematic error for statistical error by truncating \eq{alg} at increasingly large $n_{max}$. %

\Tab{table1} displays results of a simulation of  $1.25 \times 10^7$ $\phi$ configurations blocked into 250 blocks of 50,000 each, for the model \eq{model} at $\tau=1000$ and $g=1/2$.  We give the conventional estimate $\CE_\tau=-1/\tau \ln\mybar C_\tau$ and estimates based on the cumulant expansion \eq{alg} truncated at various $n_{max}$, where the exact value is $\CE=0$.
For each method we give the computed value  for $\CE$ and the statistical error; for the cumulant expansion we also give the exact systematic error from truncating \eq{alg} at $n=n_{max}$ using our analytic expressions for $\kappa_n$. These numbers show how the conventional method gives a wrong answer with deceptively small statistical error. For the cumulant expansion one sees again the trade of systematic error for statistical error with increasing $n_{max}$.  \Tab{table1} shows the combined error is minimized for $n_{max}=3$, justified by noting that the $n_{max}=4$ result with statistical errors encompasses the $n_{max}=3$ result; we suggest this as a practical algorithm for determining where to truncate the cumulant expansion in general.

\begin{figure}[t]
\centerline{\includegraphics[width=6.8 cm]{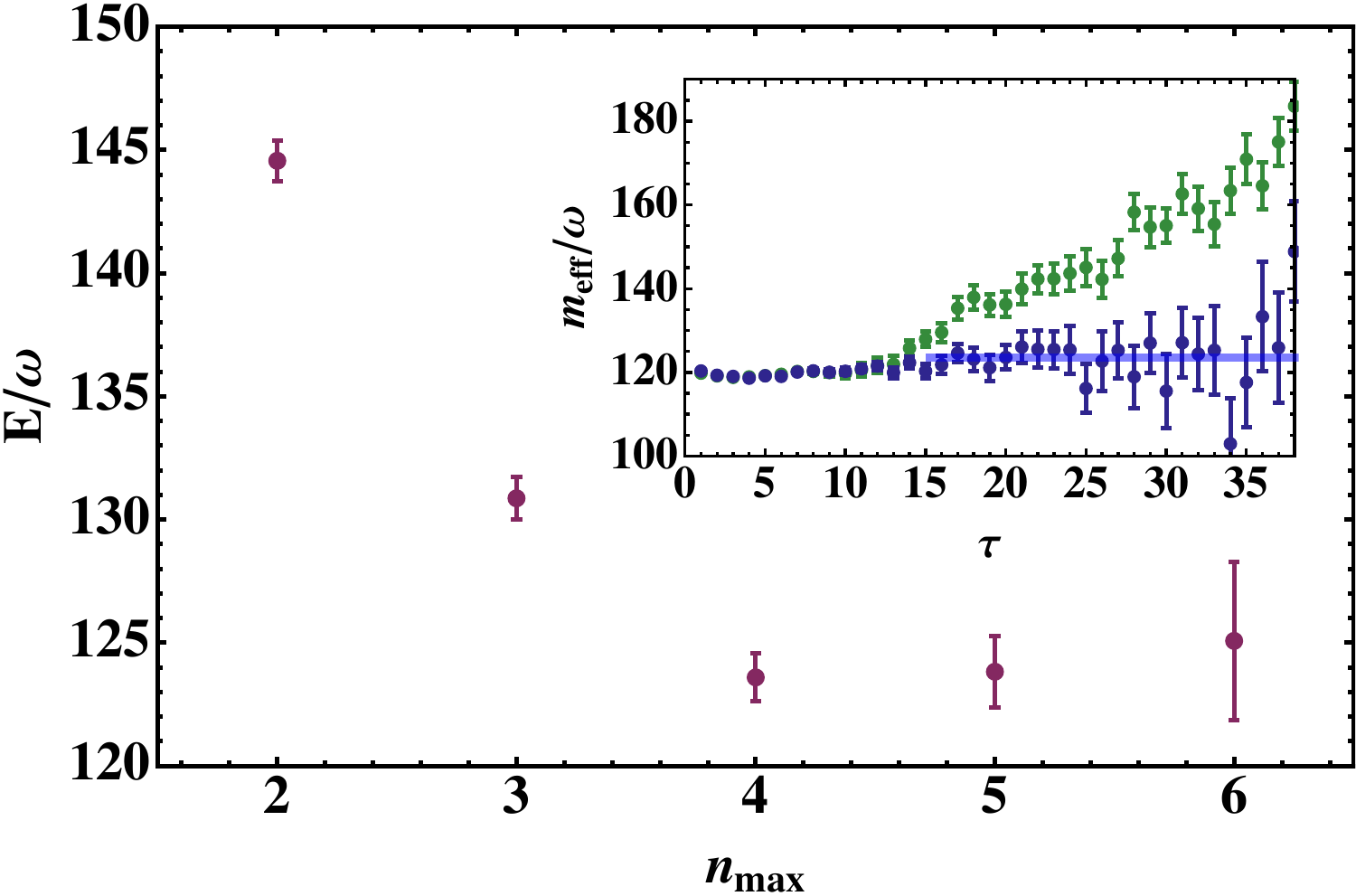}}
\caption{ Energy for 50 unitary fermions in a harmonic trap of frequency $\omega$,  $10^6$ configurations; fits performed using 
 truncated at order $n_{max}$.
Inset: conventional effective mass $m_{eff}(\tau) = \log \mybar C(\tau)/\mybar C(\tau+1)$ (green) and corresponding fitted cumulant effective mass with $n_{max}=4$ (blue).
} 
\label{fig:Convergence}
\end{figure}

\begin{figure}[t]
\centerline{\includegraphics[width=14 cm]{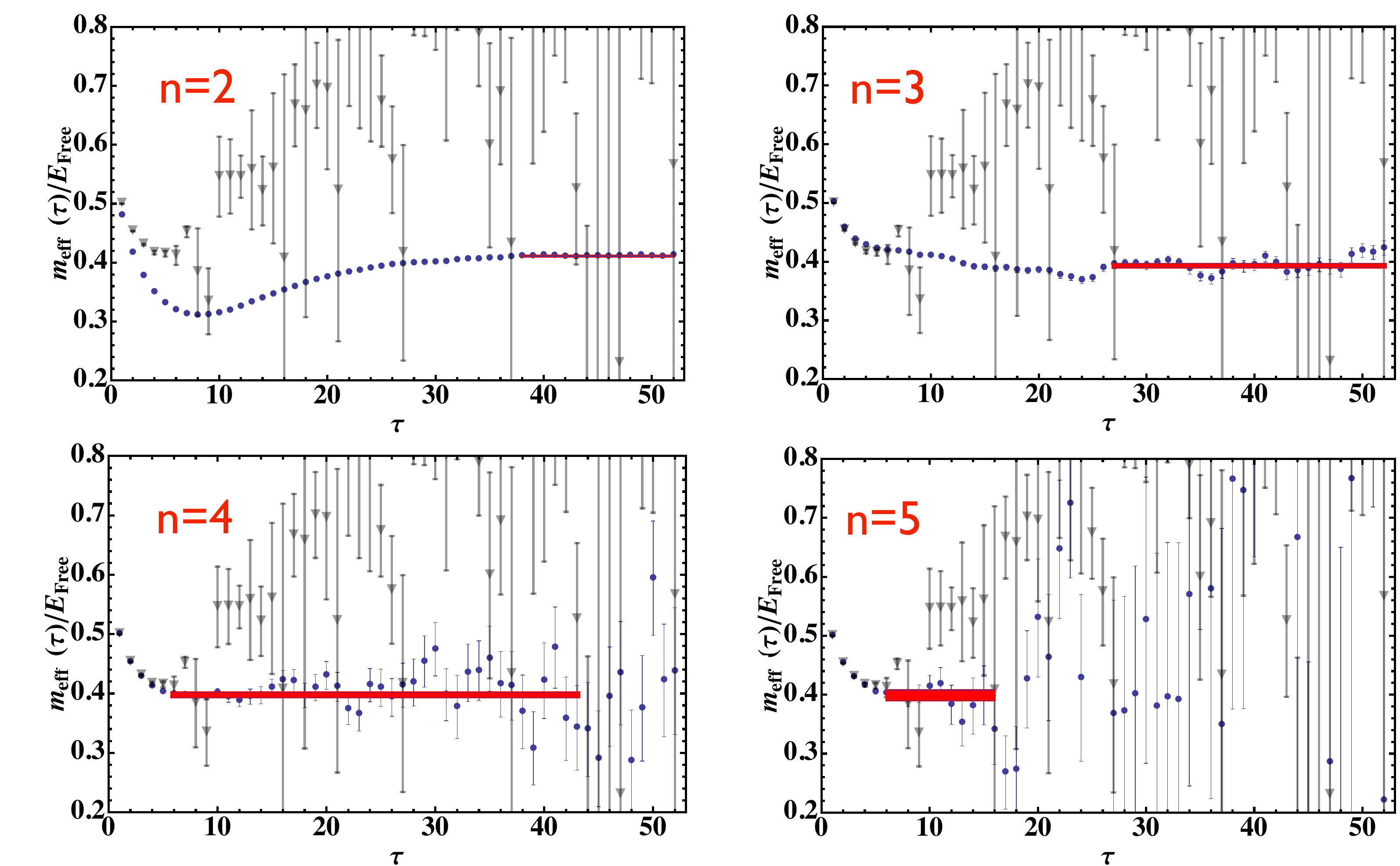}}
\caption{Effective mass plot for the same data displayed in Fig.~3, 
including $n=2,3,4,5$ terms in the expansion eq.~(4.1).  
The conventional  analysis is given by gray triangles; the cumulant expansion results  are blue circles.  In each case we have marked the plateaus we extract, with bands whose widths signify the computed error. If the correlator distribution was exactly log-normal, the $n=2$ analysis would be exact.
} 
\label{fig:noisecorrection}
\end{figure}

Leaving the toy model and returning to real simulations of unitary fermions, Fig.~\ref{fig:Convergence}  shows how the cumulant expansion works for 50 trapped unitary fermions \cite{unitary:2011aa}, where truncating the expansion at $n_{max}=4$ is supported by the data.  In Fig.~\ref{fig:noisecorrection} we show the improvement to the same data displayed above in Fig.~\ref{fig:N46noise} when we use the expansion \eq{alg} to order $n=2,3,4,5$. Again, as in the toy model, it is evident from this plot how increasing the number of cumulants included in the truncated version of \eq{alg} decreases the systematic error, but at the cost of increased statistical error.

\section{Discussion}

\begin{figure}[t]
\centerline{\includegraphics[width=12 cm]{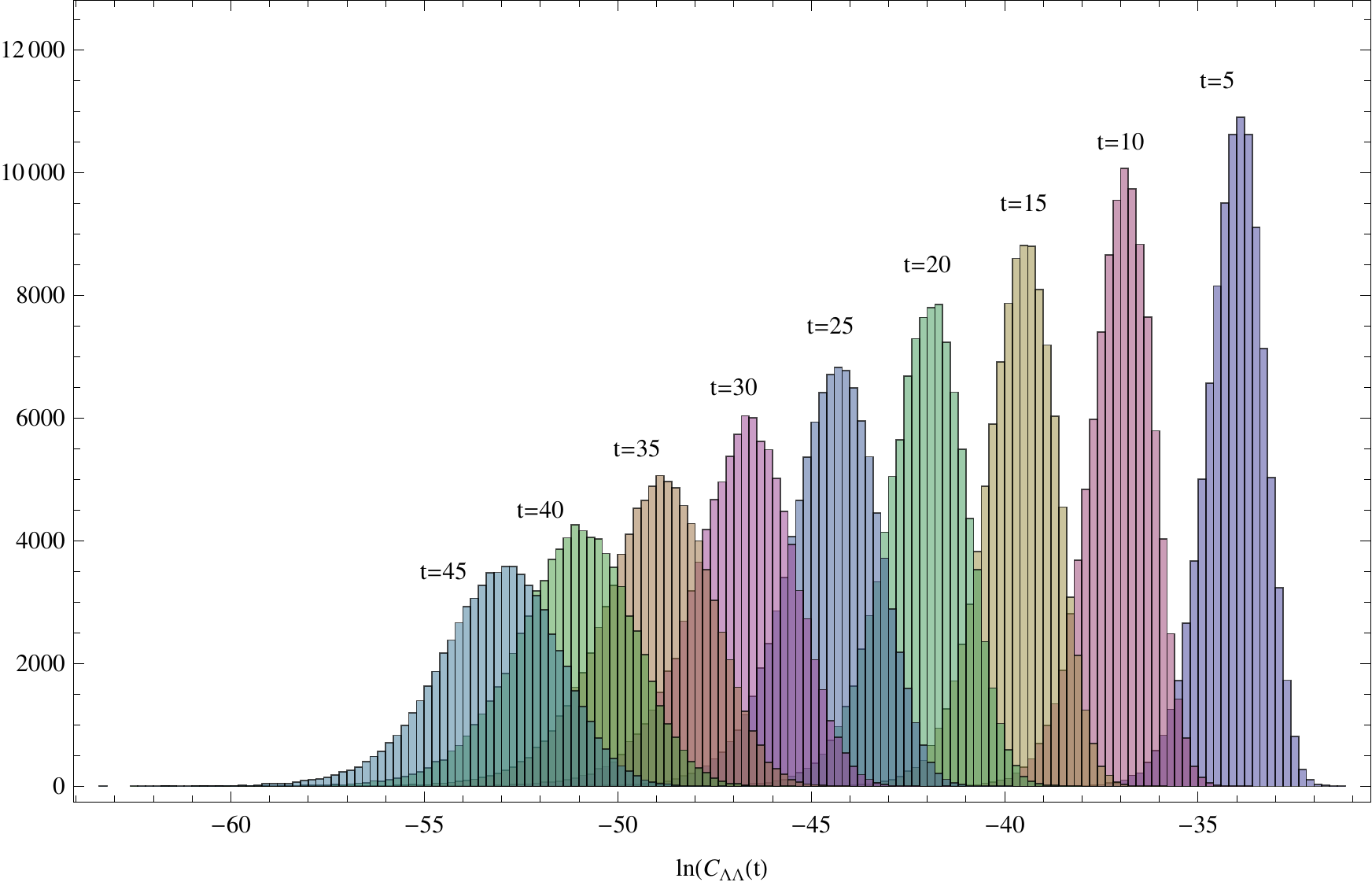}}
\caption{Statistical distribution of the log of the real part of $\Lambda\Lambda$ correlators at different temporal extent $t$ suggesting a nearly log-normal distribution for the correlators. Each curve represents 100,000 samples; a small number of correlators with negative real part were discarded. Figure from W. Detmold (NPLQCD).
} 
\label{fig:LamLam}
\end{figure}

Heavy-tail distributions  are likely to be ubiquitous in $N$-body simulations, including multi-baryon computations in lattice QCD. An example of the latter is given in Fig.~\ref{fig:LamLam}, consisting of  histograms of the log of the $\Lambda\Lambda$ correlator, as computed by the NPLQCD collaboration from  lattice QCD simulations.   This plot shows evidence that the correlator must have a nearly log-normal distribution
during the times considered.  Presumably at very late times  the distribution of the correlator will go over to a Gaussian with large variance and small mean, following the Lepage argument; however, at these earlier times that argument evidently fails, and the types of statistical techniques described here and in Ref.~\cite{endres2011noise} might improve the analysis of such QCD data.  It is interesting to speculate whether evidence for log-normal behavior in QCD implies that a mean field description could be valid for QCD correlators at moderate times, as discussed in \S\ref{sec:mf};  note the similarity between the QCD example Fig.~\ref{fig:LamLam}, and the analogous plot for unitary fermions, Fig.~\ref{fig:ln} --- as well as the different way the two distributions scale with time. One might expect that if such a mean field description is approximately valid for multibaryon correlators in QCD, as baryon number is increased that description would become even more accurate, and  log-normal behavior would become ubiquitous.

It is also interesting to wonder whether the expansion in \eq{alg} could be reformulated in an effective field theory language, where higher cumulants play the role of irrelevant operators.  The reason the correlator distribution is flowing toward a simple universal form (log-normal) is presumably because fermions propagating through a single background field configuration are sampling many values of that field which are nearly statistically independent. What one measures is a sum of products of weakly correlated random matrices.  The log of the correlator is in some sense  averaging over short range fluctuations in the field, and it must be this averaging process --- similar to RG blocking --- which accounts for the flow of the distribution toward a Gaussian infrared attractor, with deviations characterized by the cumulants of the log of the correlator. Analogies between the RG and statistics are well known --- an example being the Central Limit Theorem (see, for example, \cite{sornette2004critical}), or extreme statistics \cite{gyšrgyi2010renormalization}. If the physical basis for attractors for the statistical distribution of correlators was better understood (not much is known about the distributions of products of random matrices, let alone sums of such products, but see \cite{PhysRevE.66.066124}), including a systematic parameterization of deviations from the fixed point  --- analogous to the systematic inclusion of the effects of irrelevant operators in an effective field theory ---  then it is conceivable that one could greatly enhance one's ability to find a signal hiding in what superficially looks like useless noise.


\begin{acknowledgments}
We are greatly indebted to the NPLQCD collaboration for sharing some of their lattice QCD data with us, and in particular to W. Detmold and K. Orginos. This work was supported in part by U.S.\ DOE grant No.\ DE-FG02-00ER41132.
M.G.E is supported by the Foreign Postdoctoral Researcher program at RIKEN.
\end{acknowledgments}

\bibliography{Stats}

\end{document}